\documentclass[onecolumn,showpacs,amssymb,prd,nofootinbib]{revtex4}

\usepackage{graphicx}
\usepackage{dcolumn}
\usepackage{bm}

\def\gtap{\mathrel{ \rlap{\raise 0.511ex \hbox{$>$}}{\lower 0.511ex
   \hbox{$\sim$}}}} 
\def\ltap{\mathrel{ \rlap{\raise 0.511ex
    \hbox{$<$}}{\lower 0.511ex \hbox{$\sim$}}}} 
\newcommand{\bea}{\begin{eqnarray}} 
\newcommand{\eea}{\end{eqnarray}}
\def\beq{\begin{equation}}
\def\enq{\end{equation}}
\def\ba{\begin{eqnarray}}
\def\ea{\end{eqnarray}}
\def\<{<\!\!}
\def\>{\!\!>}
\def\<{\langle}
\def\>{\rangle}

\begin{document}

\vskip-6pt \hfill  \hspace{15truecm}
\begin{tabular}{l}
                 {IPPP/07/81} \\
                 {DCPT/07/162}
                 \end{tabular}

\input{epsf}

\vspace{0.6cm}

\title{Testing MeV dark matter with neutrino detectors}

\author{Sergio Palomares-Ruiz and Silvia Pascoli} 

\affiliation{IPPP, Department of Physics, Durham University,
  Durham DH1 3LE, United Kingdom} 

\vspace{6mm}
\renewcommand{\thefootnote}{\arabic{footnote}}
\setcounter{footnote}{0}
\setcounter{section}{0}
\setcounter{equation}{0}
\renewcommand{\theequation}{\arabic{equation}}

\begin{abstract}
MeV particles have been advocated as Dark Matter (DM) candidates in
different contexts. This hypothesis can be tested indirectly by
searching for the Standard Model (SM) products of DM
self-annihilations. As the signal from DM self-annihilations depends
on the square of the DM density, we might expect a sizable flux of
annihilation products from our galaxy. Neutrinos are the least
detectable particles in the SM and a null signal in this channel
would allow to set the most conservative bound on the total
annihilation cross section. Here, we show that neutrino detectors with
good energy resolution and low energy thresholds can not only set
bounds on the annihilation cross section but actually test the
hypothesis of the possible existence of MeV DM, i.e. test the values
of the cross section required to explain the observed DM density. At
present, the data in the (positron) energy interval [18--82]~MeV of the
Super-Kamiokande experiment is already able to put a very stringent
bound on the annihilation cross section for masses between
$\sim$15-130~MeV. Future large experiments, like megaton
water-\v{C}erenkov or large scintillator detectors, will improve the
present limits and, if MeV DM exists, would be able to detect it.
\end{abstract}

\maketitle

\section{Introduction}
\label{intro}

The existence of Dark Matter (DM) in the Universe, established from 
cosmological and astrophysical observations, remains an unsolved
puzzle. Despite the precision with which the DM abundance is
known~\cite{Spergel:2006hy}, we have still limited information on its
properties and its nature. Various heavy candidates have been
proposed, from neutralinos, sneutrinos and gravitinos in
supersymmetry, to stable scalars in little Higgs models, to
Kaluza-Klein modes in extra-dimensions and superheavy candidates (for
a review see, e.g. Ref.~\cite{Bertone:2004pz}). Light particles have
also been studied as possible DM constituents: axions~\cite{axions},
sterile neutrinos with masses in the keV range~\cite{DW92} and light
scalars with MeV--GeV
masses~\cite{Boehm:2003hm,Boehm:2003bt,Hooper:2003sh,Boehm:2006mi}.

Here, we focus on light particles as DM candidates: they can evade the
Lee-Weinberg limit~\cite{Lee:1977ua} and successfully explain the
observed amount of DM in the Universe. They can be coupled to heavy
fermions~\cite{Boehm:2003hm,Boehm:2003bt,Boehm:2006mi} or to new light 
gauge bosons~\cite{Boehm:2003hm}. For masses below 100~MeV, the only
Standard Model (SM) channels of annihilation allowed are into
electron-positron pairs, photons and neutrino-antineutrino
pairs. Interestingly, if MeV DM annihilates into SM particles other
than neutrinos, it could make an important contribution to the
reionization of the Universe, as well as significantly raise the gas
temperature prior to the reionization epoch, leaving a potentially
detectable imprint on the cosmological 21-cm
signal~\cite{Chuzhoy} (see Refs.~\cite{otherreion} for different
results). However, the couplings to electrons are strongly constrained
by observations of gamma rays. In particular, the DM
self-annihilations into electron-positron pairs is one of the
candidate explanations~\cite{Boehm:2003bt} for the 511~keV emission
line detected by Integral/SPI from the bulge of our
galaxy~\cite{511keV}. Nevertheless, the required cross section (times 
the relative velocity) is, by about four orders of magnitude, smaller
than the one necessary to explain the observed abundance of DM in the
Universe, $\langle \sigma_{\rm A} v \rangle \simeq 3 \times 10^{-26}
{\rm cm}^3/{\rm s}$. Only velocity-dependent cross sections would be
able to accommodate both evidences, but provide a bad fit to the
Integral/SPI data for masses below
$\sim$100~MeV~\cite{Ascasibar:2005rw,Rasera:2005sa}. In addition,
higher order processes would lead to gamma ray production via internal
bremsstrahlung~\cite{Beacom:2004pe,BU06,SCS06} and positron
annihilations in flight with the electrons in the interstellar
medium~\cite{Beacom:2005qv,SCS06}. The COMPTEL and EGRET data
constrain either the cross section to be smaller than what is required
to explain the 511~keV line or the DM mass to be smaller than a few
MeV. Similarly, the annihilation into photons can be constrained even 
stronger than the electron-positron channel by searches of diffuse
gamma rays. Thus, the self-annihilation cross section into
electron-positron pairs or photons, either was subdominant at DM
freeze-out in the Early Universe or, if velocity-dependent, at
present it is much smaller than it was at DM freeze-out and cannot be
tested today in a model independent way.

It is natural to assume that these light DM particles couple to
neutrinos as well, unless special models are invoked. While the
couplings with charged leptons and photons are severely bounded as 
discussed above (see also
Refs.~\cite{colLDMbounds,nuLDMbounds,atomLDMbounds,astroLDMbounds}),
the interactions with neutrinos are very poorly constrained and can be
even stronger than weak interactions. Even more, the self-annihilation
into neutrinos, $\chi \chi \rightarrow \nu \nu$, is the only channel
into SM particles, whose velocity-independent cross section can be as
large as required to reproduce the observed DM abundance! Therefore,
the analysis of constraints on the $\chi \chi \rightarrow \nu \nu$
cross section might constitute a direct test of DM freeze-out, for DM
with masses in the MeV range.

Interestingly, it was recently noted~\cite{Boehm:2006mi}
that, for MeV range DM, the value of the annihilation cross section
necessary to reproduce the observed DM abundance may induce neutrino
masses in the experimentally constrained range of values via one-loop
contributions. In addition, MeV DM could also have implications for
observations of small scale structure~\cite{Boehm,HKSZ07}. On the
other hand, it has also been noticed that DM could couple to neutrinos
strongly enough to produce observable effects that can be constrained
by cosmic microwave background~\cite{MMSCK06} and large-scale
structure formation observations~\cite{Boehm,MMSCK06}. However, in
this case, only if an asymmetry between DM particle and antiparticle
is produced at some early epoch in the evolution of the Universe, the
DM density could be the required one today. 

These models of light DM can be tested indirectly by the detection of
SM particles (electrons, photons and neutrinos) emitted in DM
self-annihilations in galaxies, in particular the Milky Way. Here, we
will assume the branching ratio into neutrinos to be dominant in DM 
self-annihilations. Since neutrinos are the least detectable particles
in the SM, a limit on their flux would conservatively set an upper
bound on the total annihilation cross section. The same approach, but
for higher masses, was followed in Refs.~\cite{BBM,YHBA}. We evaluate
the expected signal from DM annihilations in the entire Milky Way in
large neutrino detectors with low energy threshold, such as the
existing Super-Kamiokande (SK) and the proposed liquid scintillator
detector LENA~\cite{LENA}. By using SK data we are able to set a very
stringent upper bound on the total annihilation cross section in the
mass interval $\sim$15-130~MeV. In addition, for velocity-independent
cross sections and DM masses below $\sim$130~MeV, this analysis
constrains the only SM channel which could have been relevant at DM
freeze-out in the Early Universe. Finally, we also study the
possibilities to detect a positive signal from MeV DM in future
detectors like the proposed LENA detector. 

In Sec.~\ref{nuflux}, we compute the neutrino flux from DM
self-annihilations and briefly discuss halo profile uncertainties, and
in Sec.~\ref{nudet} we review the commonly used techniques to detect
MeV neutrinos and the main sources of background at these
energies. The continuum background plays an important role in the
sensitivity of present and future experiments and we discuss it in
detail. In Sec.~\ref{SKbound}, we obtain a conservative bound on the
total annihilation cross section from the present SK data, while we
devote Sec.~\ref{futuresens} to the study of the sensitivity of future
experiments and their ability to test the hypothesis of light
DM. Finally, we draw our conclusions in Sec.~\ref{conclusions}.

\section{Neutrino fluxes from Dark Matter self-annihilations}
\label{nuflux}

Stable particles with masses in the MeV range constitute a cold DM
candidate. Detailed structure formation simulations show that cold DM
clusters hierarchically in halos and the formation of large scale
structure in the Universe can be successfully reproduced. In the case
of spherically symmetric matter density with isotropic velocity
dispersion, the simulated DM profile in the galaxies can be
parametrized via
\begin{equation}
\rho(r) = \rho_{\rm sc} \,
  \left(\frac{R_{\rm sc}}{r}\right)^\gamma \, 
  \left[\frac{1+(R_{\rm sc}/r_{\rm s})^\alpha}{1+
  (r/r_{\rm s})^\alpha}\right]^{(\beta-\gamma)/\alpha},  
\label{rhopar}
\end{equation}
where $R_{\rm sc}=8.5$~kpc is the solar radius circle, $\rho_{\rm sc}$
is the DM density at $R_{\rm sc}$, $r_{\rm s}$ is the scale radius,
$\gamma$ is the inner cusp index, $\beta$ is the slope as $r
\rightarrow \infty$ and $\alpha$ determines the exact shape of the
profile in regions around $r_{\rm s}$. Commonly used profiles
~\cite{NFW,Kravtsov,Moore} (see also Ref.~\cite{DMprofiles}) can
differ considerably in the inner part of the galaxy. However, this has
only a relatively mild effect on the neutrino flux if a large field of
view is considered~\cite{YHBA}.
 
Assuming DM annihilates into neutrino-antineutrino pairs, we can
compute the neutrino flux coming from these annihilations in the halo 
(see, e.g. Ref.~\cite{BUB97}). As we will see below, for energies
below $\sim$ 100~MeV, information on the direction of the incoming
neutrino is very poor if the detection is via interactions with
nucleons. As this is the case presented here, we will be interested in
considering the flux coming from all directions, so we shall take the
flux averaged over the entire galaxy. This also helps in maximizing
the observed neutrino flux while minimizing the impact of the choice
of DM profile~\cite{YHBA}. 

The angular-averaged intensity over the whole Milky Way, i.e. the
average number flux, is given by the angular-averaged line of sight
integration of the DM density square,
\begin{equation}
{\cal J}_{avg}  = \frac{1}{2R_{\rm sc} \, \rho^2_{0}}  \,
 \int_{-1}^1 \, \int_{0}^{l_{\rm max}} \,  \rho^2 (r) \, dl \, d(\cos
 \psi),
\label{Javg}
\end{equation}
where $r = \sqrt{R^2_{\rm sc} -  2 l R_{\rm sc} \cos \psi + l^2}$,
$\rho_0 = $ 0.3~GeV~cm$^{-3}$ is a normalizing DM density, which is
equal to the commonly quoted DM density at $R_{\rm sc}$, and the upper
limit of integration is $l_{\rm max} = \sqrt{(R_{\rm halo}^2 - \sin^2
\psi R^2_{\rm sc})} + R_{\rm sc} \cos \psi$, and depends on the size
of the halo $R_{\rm halo}$. As the contribution at large scales is
negligible, different choices of $R_{\rm halo}$ do not affect $ {\cal
J}_{avg}$ in a significant way, as long as it is a factor of a few
larger than the scale radius, $r_{\rm s}$.
 
The differential neutrino and antineutrino flux per flavor from DM
annihilations is then given by
\begin{equation}
\frac{ d \phi}{d E_\nu} = \frac{\langle \sigma_{\rm A} v \rangle}{2}
 {\cal J}_{avg} \, \frac{R_{\rm sc} \, \rho_{0}^2}{m_\chi^2} \,
 \frac{1}{3} \, \delta(E_\nu - m_\chi),
\label{diffflux}
\end{equation}
where $m_\chi$ is the mass of the DM particle and $\langle \sigma_
{\rm A}  v \rangle$ is the averaged self-annihilation cross
section (times the relative velocity of the annihilating
particles). The factor 1/2 accounts for DM being its own antiparticle,
and the factor of 1/3 comes from the assumption that the branching 
ratio of annihilation is the same in the three neutrino flavors. In
case of annihilation predominantly in one flavor, averaged oscillations
between the production point and the detector would generate the other
flavors with comparable intensity. In particular, given the present
oscillation parameters~\cite{Maltoni:2004ei}, for pure $\nu_\mu$ and
$\nu_\tau$ channels, the final flavor ratios would be
$\nu_e:\nu_\mu:\nu_\tau \simeq$  1 : 2 : 2, whereas for a pure $\nu_e$
flux we would have $\nu_e:\nu_\mu:\nu_\tau \simeq$  3 : 1 : 1. Hence,
thanks to neutrino oscillations, there is a guaranteed flux of
neutrinos in all flavors. However, for simplicity herein we will
consider equal annihilation into all flavors. 

Let us note that while DM profiles tend to agree at large scales,
uncertainties are still present for the inner region of the galaxy. As
the neutrino flux from DM annihilations scales as $\rho^2$, this leads
to an uncertainty in the overall normalization of the flux. To
understand this effect quantitatively, we have studied the impact of
the chosen halo profile by choosing three spherically symmetric
profiles with isotropic velocity dispersion which, from more to less
cuspy, are: Moore, Quinn, Governato, Stadel and Lake
(MQGSL)~\cite{Moore}, Navarro, Frenk and White (NFW)~\cite{NFW} and
Kravstov, Klypin, Bullock and Primack (KKBP)~\cite{Kravtsov}. For each
of the three profiles there is a range of values for the DM density at
the solar circle, $\rho_{\mathrm{sc}}$, which satisfy the present
constraints from the allowed range for the local rotational
velocity~\cite{v0}, the amount of flatness of the rotational curve of
the Milky Way and the maximal amount of its non-halo
components~\cite{massMW}. We compile in Table~\ref{table} the values
of the parameters in Eq.~(\ref{rhopar}), the limiting values for
$\rho_{\rm sc}$~\cite{BCFS02}, along with the corresponding limiting
values of ${\cal J}_{avg}$ for each of the three profiles. Note also
that, by choosing the same $\rho_0$ for all profiles, all the
uncertainties in the neutrino flux from DM annihilations, coming from
the lack of knowledge of the halo profile, lie in ${\cal J}_{avg}$.

Finally, let us mention that the diffuse signal from cosmic
annihilations from all halos could in principle also be used to
constrain the total DM annihilation cross section in the mass range
considered here. However, the formation history of halos, although
with a fairly universal dependence on the redshift, has a normalization
which is quite uncertain and varies by several orders of magnitude
for different halo profiles~\cite{Ando05}. In addition to the
uncertainties of the cosmic signal induced by those in the halo
profiles, other important factors as the clustering of halos, the
halo mass function and the lower mass cutoff, are not completely well
known and introduce further uncertainties in the cosmic
signal. Therefore, the cosmic signal, being much more uncertain than
that from the whole Milky Way, is less suitable to obtain reliable
bounds on the total DM annihilation cross section. In addition, the
cosmic signal is likely to be smaller than, or at most of the same
order of, the galactic one~\cite{YHBA}. Hence, and although the
diffuse (redshifted) cosmic signal would allow to obtain bounds on the
total DM annihilation cross section for higher masses than the
galactic signal studied here, we will not consider it further.

\begin{table}[t]
\caption{The parameters of  Eq.~(\ref{rhopar}) for the three halo
  profiles considered. Also shown the limiting values of $\rho_{\rm
  sc}$ in [GeV cm$^{-3}$] for the three profiles, as calculated in
  Ref.~\cite{BCFS02}, which satisfy the present constraints from the 
  allowed range for the local rotational velocity~\cite{v0}, the
  amount of flatness of the rotational curve of the Milky Way and the
  maximal amount of its non-halo components~\cite{massMW}. Also shown
  the corresponding limiting values for ${\cal J}_{avg}$ which
  directly affect the total flux of the DM annihilation products. We
  will also use the canonical value of Ref.~\cite{YHBA}, ${\cal
  J}_{avg}$ = 5. To avoid numerical divergences, when computing ${\cal
  J}_{avg}$, we consider a flat core in the innermost 0.1$^o$.\\}
\label{table} 
\begin{tabular}{l|cccccccc} 
\hline
\hline \\
 \hfill & $\alpha$ \hfill& $\beta$ \hfill & $\gamma$ \hfill & $r_{\rm
   s}$ [kpc] \hfill & $(\rho_{\rm sc})_{\rm min}$ [GeV/cm$^3$] \hfill
 & $(\rho_{\rm sc})_{\rm max}$ [GeV/cm$^3$] \hfill & $({\cal
   J}_{avg})_{\rm min}$ \hfill & $({\cal J}_{avg})_{\rm max}$ \hfill
 \\[1ex] \hline \\ 
MQGSL~\cite{Moore} \hfill & 1.5 \hfill & 3 \hfill & 1.5 \hfill & 28
\hfill & 0.22 \hfill & 0.98 \hfill & 5.2 \hfill & 104 \hfill \\[1ex] 
NFW~\cite{NFW} \hfill & 1 \hfill & 3 \hfill & 1 \hfill & 20 \hfill &
0.20 \hfill & 1.11 \hfill & 1.3 \hfill & 41 \hfill \\[1ex] 
KKBP~\cite{Kravtsov} \hfill & 2 \hfill & 3 \hfill & 0.4 \hfill & 10
\hfill & 0.32 \hfill & 1.37 \hfill & 1.9 \hfill & 8.5 \hfill \\[1ex]
\hline
\hline
\end{tabular}
\end{table}

\section{M\lowercase{e}V neutrino detection}
\label{nudet}
 
The neutrinos produced in DM self-annihilations travel from their
production point in the galaxy to the Earth where they can be revealed
in present and future neutrino detectors. The number of neutrino
events in a given detector is given by
\begin{equation}
{\cal N} \simeq \sigma_{\rm det}(m_\chi) \  \phi \
N_{\rm target} \ t  \, \epsilon   ~,
\label{nevents}
\end{equation}
where the detection cross section $\sigma_{\rm det}$ needs to be
evaluated at $E_\nu = m_\chi$, the total flux of neutrinos (or
antineutrinos) is given by $\phi$, $N_{\rm target}$ indicates the
number of target particles in the detector, $t$ is the total 
time-exposure, and $\epsilon$ is the detector efficiency for this
type of signal. 

At the energies of interest, in the range of tens of MeV, the inverse
beta-decay cross section ($\overline{\nu}_e p\rightarrow n e^+$) is by
two orders of magnitude larger that the $\nu-e$ elastic scattering
cross section. Neutrino interactions with free protons are also
stronger than interactions with bound nucleons up to energies of about
$\sim$80~MeV. However, the latter interactions give non-negligible
contributions and should be taken into account. Thus, in the
following, we focus on $\overline{\nu}_e$ from DM annihilations and on
neutrino interactions on both free and bound nucleons. However, for
detectors with very low energy threshold like scintillator detectors,
the inverse beta-decay reaction can be clearly tagged by the signal in
coincidence of the positron annihilation followed by a delayed 2.2~MeV
photon, which is emitted when the neutron is captured by a free
proton. In water-\v{C}erenkov detectors like SK, the threshold is
above 2.2~MeV, so this reaction cannot be discriminated from neutrino
interactions with nuclei. If the detector is doped with gadolinium
trichloride (GdCl$_3$)~\cite{GADZOOKS!}, the neutron capture on Gd
leads to 3--4 photons with a total energy of 8~MeV, and helps in this
discrimination.

From Eq.~(\ref{nevents}), and assuming the annihilation cross section
required to reproduce the observed amount of dark matter, $\langle
\sigma_{\rm A} v \rangle \simeq 3 \times 10^{-26} \, {\rm cm}^3/{\rm
s}$, we expect few events for an exposure of a Mton $\cdot$
yr. Therefore, we restrict our analysis to large neutrino detectors
with low energy thresholds as SK and the proposed LENA scintillator
detector~\cite{LENA}. Similar considerations would apply also for
proposed Liquid Argon detectors as GLACIER~\cite{GLACIER}.

If the detection technique allows to distinguish the inverse beta-decay
reaction from neutrino interactions off nuclei (either with a
scintillator detector or with a water-\v{C}erenkov detector doped with
GdCl$_3$), the expected signal has a very specific experimental
signature being given by a peak (sharper the lower the DM mass is) in
the neutrino spectrum and would be easily distinguished from the
continuum background if a sufficient energy resolution is
available. In general, the backgrounds for these events are due to
geoneutrinos, solar and reactor  neutrinos, to muon-induced spallation
products, and to atmospheric and diffuse supernova neutrinos: 

i) {\it Geoneutrinos}, produced in the Earth interior, result from
the radioactive decay chains of nuclear isotopes with lifetimes
comparable to or longer than the Earth age, like $^{238}$U,
$^{232}$Th, $^{40}$K, $^{235}$U and $^{87}$Rb (see,
eg. Ref.~\cite{FLM07}). Their spectrum extends up to 3.27~MeV with a 
total flux, $\sim 10^6 \, \overline{\nu}_e \, {\rm
cm}^{-2} \, {\rm s}^{-1}$, several orders of magnitude higher than
that of neutrinos from DM self-annihilations being larger in regions
with a thick continental crust.  

ii) {\it The solar neutrino flux}, similar in magnitude to that of
geoneutrinos, drops rapidly at energies $\gtap$10~MeV, and in
water-\v{C}erenkov detectors it can be eliminated by applying an
angular cut by exploiting the directionality of $\nu - e$ elastic
scattering. On the other hand, the solar neutrino flux being a flux of
neutrinos, in scintillator detectors it induces a different type of
signal from inverse beta-decay and does not constitute a relevant
background.

iii) {\it The flux of reactor $\overline{\nu}_e$'s}, below 10~MeV,
is by orders of magnitude higher than the expected neutrino flux from
DM self-annihilations. In reactors, $\overline{\nu}_e$ are generated
in the beta-decay of the fission products of $^{235}$U, $^{238}$U,
$^{239}$P and $^{241}$P. The shape and the normalization flux can be
estimated with good precision taking into account experimental data
and theoretical calculations. The tail of the spectrum, up to
$E_{\overline{\nu}_e} = 13 \, {\rm MeV}$, is less well known due to
the presence of $^{94}$Br with Q-value of 13.3~MeV, whose exact decay
scheme is not know. However, its continuum spectrum could be easily
distinguished from a peaked neutrino signal from DM
annihilation. Typically, one expects a flux of $\sim 10^6 \,
\overline{\nu}_e \, {\rm cm}^{-2}\, {\rm s}^{-1}$ at sites as Kamioka 
(Japan) and Frejus (France), and an order of magnitude smaller at
Pyh\"asalmi (Finland), Homestake (US) and Henderson (US). The lowest
reactor fluxes can be found at Hawaii (US), $1 \times 10^{4} \, {\rm
cm}^{-2} {\rm s}^{-1}$, and Wellington in New Zealand, $5 \times
10^{3} \, {\rm cm}^{-2} {\rm s}^{-1}$, due to their distance from the
northern hemisphere power plants~\cite{LENADSNB}. Even in the latter
locations, the neutrino flux from DM annihilations is by orders of
magnitude smaller. Thus, we restrict the analysis to energies above 10
MeV.

iv) {\it Muon-induced spallation products} constitute a very important
background in water-\v{C}erenkov detectors and a sufficient reduction
sets the lower energy threshold of the detector. For SK, a threshold
of 18~MeV has been set and a tight spallation cut used. An efficient
background reduction can be obtained in this way~\cite{SKSN}. In
scintillator detectors, as LENA, the topology of the event allows to
reduce this background sufficiently~\cite{LENADSNB}.

v) {\it The flux of atmospheric $\nu_e$ and $\overline{\nu}_e$}
should also be taken into account in the energy window under
consideration. The normalization of the flux depends on the location
of the detector~\cite{Bartol,HKKM95,LDB03,FLUKA}, more
specifically on the geomagnetic latitude, varying roughly by a factor
of 2 between Hawaii (1.5$^o$ latitude N) and Pyh\"asalmi (63.7$^o$
latitude N)~\cite{LENADSNB}. The shape of the spectrum of atmospheric
neutrinos, however, is not sensibly affected by the position of the
underground laboratory. In detectors able to tag the neutron produced
in the inverse beta-decay reaction, only the flux of
$\overline{\nu}_e$ constitutes a relevant background.

vi) {\it Invisible muons from atmospheric $\nu_\mu$ and
$\overline{\nu}_\mu$} constitute the dominant background in 
water-\v{C}erenkov detectors. If the kinetic energy of the produced
muon is below 54~MeV, then the muon is below the threshold for
emitting \v{C}erenkov radiation. These muons, produced by atmospheric
$\nu_\mu$ and $\overline{\nu}_\mu$ with typical energies of about
$\sim$200~MeV, are slowed down rapidly and subsequently decay, giving
rise to a signal which cannot be distinguished from that of an $\nu_e$
or $\overline{\nu}_e$, and hence pose an important source of
background. On the other hand, in detectors for which it is also  
possible to tag the neutron from the inverse beta-decay, like
scintillator detectors as LENA~\cite{LENA}, these muons do not
constitute a background. In water-\v{C}erenkov detectors doped with
GdCl$_3$~\cite{GADZOOKS!}, this background can be reduced by a factor
of $\sim 5$ by rejecting events with a preceding nuclear gamma or
which are not followed by a neutron.

vii) {\it The Diffuse Supernova Neutrino Background (DSNB)}, although
not yet detected, might potentially represent a background for the
signal coming from neutrinos from DM self-annihilations. This flux
might be relevant in the interval of energies between
$\sim$10~MeV and $\sim$30~MeV, dropping very rapidly with energy, and
being completely negligible above 50~MeV--60~MeV.

\section{Bounds from SK data}
\label{SKbound}

In this section we extend previous
works~\cite{unitarity,KKT,BBM,KS,YHBA} by setting neutrino constraints 
on the dark matter total annihilation cross section for masses below
$\sim$130~MeV. In doing this we adopt a similar approach to that in
Refs.~\cite{BBM,YHBA}, and assume that dark matter annihilates only
into neutrinos. If dark matter annihilates into SM particles,
neutrinos (and antineutrinos) are the least detectable particles. Any
other possible decay mode would produce gamma rays, which are much
easier to detect, and would allow to set a much stronger (and
model-dependent) bound on the total annihilation cross section. Thus,
the most conservative approach~\cite{BBM,YHBA} is to assume that only
neutrinos are produced in DM self-annihilations. Even in this
conservative case, it has been shown that a stringent upper limit on
the total DM annihilation cross section can be obtained for masses
above $\sim$100~MeV by comparing the expected time-integrated
annihilation signal of all galactic halos (cosmic signal)~\cite{BBM}
and the signal from annihilations in the Milky Way Halo~\cite{YHBA} 
with the dominant background at these energies: the flux of
atmospheric neutrinos.

In this paper, we consider the low mass region below $\sim$130~MeV,
for which the best present data comes from the SK
detector~\cite{SKSN}: a search for the DSNB was performed by looking
for positrons in the energy range 18~MeV--82~MeV, produced by 
$\overline{\nu}_e$ interactions. The same data can be analyzed to
search for neutrinos from DM self-annihilations. As explained above,
in this energy range, the two dominant backgrounds in
water-\v{C}erenkov detectors are the atmospheric $\nu_e$ and
$\overline{\nu}_e$ flux and, mainly, the Michel electrons (positrons)
from the decays of low energy muons which are below detection
threshold. Below 18~MeV, muon-induced spallation products are the most
serious background, and it is indeed the ability to remove this
background what determines the lower energy threshold for the DSNB
search.

As mentioned above, below $\sim$80~MeV, the dominant interaction of
$\overline{\nu}_e$ is through the inverse beta-decay reaction,
$\overline{\nu}_e + p \rightarrow e^+ + n$. However, the interactions
of neutrinos with oxygen nuclei need also to be taken into account.  We
have included in our analysis both the interactions of antineutrinos
with free protons and the interactions of neutrinos and antineutrinos
with bound nucleons, by considering, in the latter case, a relativistic
Fermi gas model~\cite{SM72} with a Fermi surface momentum of 225~MeV
and a binding energy of 27~MeV. 

At very low energies, the inverse beta-decay reaction relates the
energy of the outgoing positron to that of the incoming neutrino ($E_e
\simeq E_{\nu} - 1.3$~MeV). However, at higher energies corrections of
the order of ${\cal O}(E_\nu/{\rm M})$, where M is the nucleon mass,
start to become important and the difference between the maximum and
minimum positron energy is to first order given by $\Delta E_e \approx
2 E_\nu^2/{\rm M}$. Hence, although the DM self-annihilation gives
rise to monochromatic neutrinos, the signal would have a significant
energy spread for DM masses of tens of MeV. In addition, the finite
energy resolution of the detector will also contribute to the spread
of the signal. In order to take this into account, we consider the
energy resolution of the SK I detector, which is determined by the
photocathode coverage, as given by the LINAC calibration~\cite{Malek}
up to 16.09~MeV and perform a simple fit, extrapolating to higher
energies. The energy resolution that we consider is
\begin{equation}
\sigma = 0.40 \, {\rm MeV} \, \sqrt{E/{\rm MeV}} + 0.03 \, E  ~.
\end{equation}
The monochromatic neutrino flux is folded with a gaussian energy
resolution function of width $\sigma$, $R(E_e,E_{\rm vis})$, with
$E_e$ and $E_{\rm vis}$ being the original and detected electron (or
positron) energy, respectively. We have also taken into account the
energy-dependent efficiency after all cuts, which is $\epsilon =$ 0.47
for $E_{\rm vis} < 34$~MeV and $\epsilon =$ 0.79 for $E_{\rm vis} >
34$~MeV~\cite{SKSN}.

The expected fraction of signal from DM self-annihilation into
neutrino-antineutrino pairs in the visible (positron) energy interval
$E_{\rm vis} = [E_l, E_{l+1}]$
is given by
\begin{equation}
A_l  =  A_s \, \int \left[\frac{d\sigma_{\rm
 f}^{\overline{\nu}}}{dE_e} (m_{\chi},E_e) + \frac{1}{2} \,
 \left(\frac{d\sigma_{\rm b}^{\nu}}{dE_e} (m_{\chi},E_e) +
 \frac{d\sigma_{\rm b}^{\overline{\nu}}}{dE_e} (m_{\chi},E_e)
 \right) \right] \, dE_e \times \int_{E_l}^{E_{l+1}} \, \epsilon
 (E_{\rm vis}) \, R(E_e,E_{\rm vis}) \, dE_{\rm vis}  ~,
\end{equation} 
with $E_1 = 18$~MeV and $E_{l+1} - E_l = 4$~MeV. $A_s$ is a
normalization constant so that $\sum A_l = 1$. We indicate with
$\sigma_{\rm f}$ and $\sigma_{\rm b}$ the neutrino cross sections off
free nucleons and off nuclei (bound nucleons), respectively, and the
factor of $1/2$ is due to water having twice as many free protons as
oxygen nuclei.

In order to obtain the upper limit on the total DM annihilation cross
section, we use the data reported by the SK collaboration~\cite{SKSN}
and perform an analogous analysis. We consider the sixteen 4-MeV bins
in which the data were divided and define the following $\chi^2$
function~\cite{SKSN}
\begin{equation}
\chi^2 = \sum_{l=1}^{16} \, \frac{\left[(\alpha \cdot A_l) + (\beta
    \cdot B_l) + (\gamma \cdot C_l) - N_l \right]^2}{\sigma_{stat}^2 +
    \sigma_{sys}^2} ~,
\end{equation}  
where the sum $l$ is over all energy bins, $N_l$ is the number of
events in the $l$th bin, and $A_l$, $B_l$ and $C_l$ are the fractions
of the DM self-annihilation signal, Michel electron (positron) and
atmospheric $\nu_e$ and $\overline{\nu}_e$ spectra that are in the
$l$th bin, respectively. The fractions $A_l$ are calculated as
described above. The fractions $B_l$ are calculated taking into
consideration that in water 18.4\% of the $\mu^-$ produced below
\v{C}erenkov threshold ($p_\mu < 120$~MeV) get trapped and enter a
K-shell orbit around the oxygen nucleus and thus, the electron
spectrum from the decay is slightly distorted with respect to the
well-known Michel spectrum~\cite{HVRA74}. In the calculation of the
fractions $B_l$ and $C_l$ we have used the low energy atmospheric
neutrino flux calculation with FLUKA~\cite{FLUKA}. Note that, in a
two-neutrino approximation and for energies below $\sim$300~MeV (where
most of the background comes from), half of the $\nu_\mu$ have
oscillated to $\nu_\tau$, whereas $\nu_e$ remain
unoscillated. Although this approximation is not appropriate, in
principle, to calculate the low energy atmospheric neutrino
background, however, for practical purposes, it introduces very small
corrections~\cite{FLMM04}. Thus, in order to calculate $B_l$ and $C_l$
we use the two-neutrino approximation.

The fitting parameters in the $\chi^2$-function are $\alpha$, $\beta$
and $\gamma$, which represent the total number of each type of
event. For the systematic error we take $\sigma_{sys} = 6 \%$ for all
energy bins~\cite{SKSN}. Our fit agrees very well with the
SK analysis, and we obtain as best fit points, $\beta_{\rm BF}=181 \pm
23$ and $\gamma_{\rm BF}=80 \pm 17$, with $\chi^2 = 7.9$ for 13
degrees of freedom. Our best fit to $\alpha$ is $\alpha_{\rm BF} = 0$,
which means that there are no events from DM self-annihilation into
neutrino-antineutrino pairs in the current SK data.

\begin{figure}[t]
\centerline{\epsfxsize=5.in \epsfbox{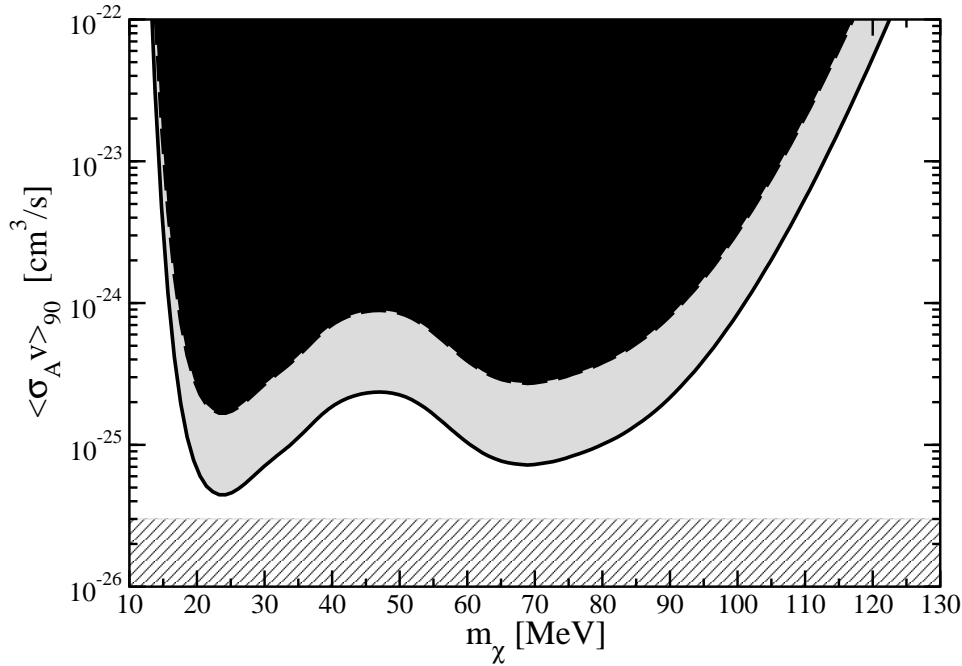}}
\caption{90\% C.L. bound on the total DM self-annihilation cross
  section from the whole Milky Way, obtained from SK data. The dashed
  line (boundary for the black area) represents the most conservative
  bound by using the smallest value observationally allowed of ${\cal
  J}_{avg}$ = 1.3, whereas for the solid line (boundary for the gray
  area), we have used the canonical value, ${\cal J}_{avg}$ = 5,
  of Ref.~\cite{YHBA}. The hatched area represents the natural scale
  of the annihilation cross section.}
\label{90CL}    
\end{figure} 

In absence of a DM signal, a 90\% confidence level (C.L.) limit can be
set on $\alpha$ for each value of the DM mass. The limit is obtained
by increasing the value of $\alpha$ and evaluating the $\chi^2$
function using $\beta$ and $\gamma$ as free parameters. The minimum
$\chi^2$ value obtained for each $\alpha$ ($\chi_{\alpha}^2$)
corresponds to a relative probability given by
\begin{equation}
P(\alpha) = K \cdot e^{-\chi_{\alpha}^2} ~,
\end{equation}
with $K$ a normalizing constant so that $\sum_{\alpha=0}^{\infty}
P(\alpha) = 1$. 

The 90\% C.L. upper limit on the number of events coming from dark
matter annihilation into neutrino-antineutrino pairs, that we also
call $\alpha_{90}$, is defined by
\begin{equation}
\sum_{\alpha=0}^{\alpha_{90}}
P(\alpha) = 0.9 ~.
\end{equation}

The limit on $\alpha$ can be translated into a limit on the
annihilation cross section, which is given by
\begin{equation}
\langle \sigma_{\rm A} v \rangle_{90} = \alpha_{90} \,
\frac{6 \, m_{\chi}^2 \, \cdot A_s}{t \cdot N_{\rm target} \cdot
  {\cal J}_{avg} \, R_{\rm sc} \, \rho_0^2}
\end{equation}
where t = 1496 days and $N_{\rm target} = 1.5 \times 10^{33}$
free protons in the fiducial volume (22.5~kton) of the SK detector. 

This limit is depicted in Fig.~\ref{90CL}. The solid line is obtained
for the canonical value of Ref.~\cite{YHBA}, ${\cal J}_{avg} = 5$,
while for the dashed line we use ${\cal J}_{avg} = 1.3$ (see
Table~\ref{table}), which represents the lowest possible value of
${\cal J}_{avg}$ for the three profiles considered. This smallest
value of ${\cal J}_{avg}$ corresponds to the NFW profile and to the
minimum density allowed for this profile by observational constraints,
$(\rho_{\rm sc})_{\rm min} = 0.2 \, {\rm GeV
  cm}^{-3}$~\cite{BCFS02}. Thus, the gray and black areas represent,
respectively, the excluded regions for the case of the canonical
value of Ref.~\cite{YHBA} and for the case of the lowest value of
${\cal J}_{avg}$. Hence, the black region is the most reliable and
conservative 90\% C.L. upper limit on the total DM annihilation
cross section we obtain. In Fig.~\ref{90CL}, the hatched region
represents the natural scale for the DM annihilation cross
section. For masses below $\sim$100~MeV, the bound is very stringent,
being within an order of magnitude of the natural scale for the most
conservative of the DM profiles, and within a factor of a few for the
canonical value of Ref.~\cite{YHBA}. Notice that more cuspy DM
profiles than the NFW profile or large values for the density at the
solar circle, $\rho_{\rm sc}$, would set a bound on the total DM
annihilation cross section below the natural scale, and hence,
excluding in those cases a velocity-independent cross section. Also
note that our bound extends beyond the interval $m_\chi = (19.3,\,
83.3)$~MeV, which would represent the na\"{\i}ve case where only the
inverse beta-decay reaction (interaction of antineutrinos off free
protons) is considered with the approximation $E_\nu = E_e + 1.3$~MeV,
and without taking into account the finite energy resolution of the
detector.

\section{Future perspectives}
\label{futuresens}

Let us now consider the possibility to detect MeV DM annihilating into
neutrino-antineutrino pairs with future neutrino detectors. We have
seen in the previous section that the bounds already set by SK data
are very close to the natural scale for masses between 10~MeV and
100~MeV, and thus, the detection (or exclusion) of this type of signal
might be possible in a matter of just a few more years. Moreover, the
addition of GdCl$_3$ to SK would allow to reduce the backgrounds by a
factor of about 5~\cite{GADZOOKS!}, which would accelerate this
search. In addition to this possibility, and also to the possible use
of larger water-\v{C}erenkov detectors (doped with GdCl$_3$ or
not)~\cite{megaton}, there are other type of large detectors which
have been proposed in the context of measuring MeV neutrinos, for DSNB
searches and for the study of solar neutrinos and geoneutrinos. We
will consider here the proposed large-volume liquid scintillator
detector LENA~\cite{LENA}, which, as noted above, will provide a very
good background discrimination.

The LENA detector is foreseen to have a fiducial volume of 50~$\times
10^3$ m$^3$ of liquid scintillator and, at the moment, one of the
preferred detector sites is a mine in Pyh\"asalmi (Finland) (another
option is underwater in the Mediterranean sea next to Pylos,
Greece)~\cite{LENA}. As mentioned above, the inverse beta-decay
reaction has a very clear signature in this type of detector, with the
positron annihilation followed by a 2.2~MeV photon, released when the
neutron is captured by a free proton in the scintillator $\sim$180
$\mu$s after the interaction takes place. In this type of detector
there is no background due to sub-\v{C}erenkov muons. In addition,
muon-induced spallation products can be efficiently removed, as well
as the background of fast neutrons generated by muons passing the
surrounding rock~\cite{LENADSNB}. Hence, the only relevant backgrounds
for the signal studied here come from reactor, atmospheric and diffuse
supernova $\overline{\nu}_e$ interacting with free protons in the
detector. The latter, albeit not yet measured, constitutes a potential
background for the signal of MeV DM annihilation into
neutrino-antineutrino pairs, but with a very different spectrum.

Below $\sim$10~MeV the flux of $\overline{\nu}_e$ from nuclear
reactors is the dominant one in LENA, and we use the reactor
$\overline{\nu}_e$ spectrum calculated in Ref.~\cite{LENADSNB}. In the
energy interval between $\sim$10~MeV and $\sim$30~MeV, the DSNB is
likely to dominate. The differential number flux of the DSNB is given
by
\begin{equation}
\label{SNspec}
\frac{dF_\nu}{dE_\nu} (E_\nu) = c \int_0^{z_{max}} R_{\rm SN}
(z) \frac{dN_\nu (E_\nu')}{dE_\nu'} (1+z) \frac{dt}{dz} dz ~,
\end{equation}
where we assume that the gravitational collapse begun at $z_{max}
=6$. The number spectrum of neutrinos emitted by one supernova is
$dN_\nu/dE_\nu$, $E_\nu' = (1+z) \, E_\nu$ is the energy of neutrinos
at redshift $z$ with $E_\nu$ being the energy at the Earth, $R_{\rm
SN} (z)$ represents the supernova rate per comoving volume at redshift
$z$ and $dt/dz = \left( H_0 (1+z) \sqrt{\Omega_M (1+z)^3 +
  \Omega_\Lambda} \right)^{-1}$. We will adopt the standard
$\Lambda$CDM cosmology ($\Omega_m = 0.3$, $\Omega_\Lambda = 0.7$ and 
$H_0 = 70$~km~s$^{-1}$~Mpc$^{-1}$).

\begin{figure}[t]
\centerline{\epsfxsize=5.in \epsfbox{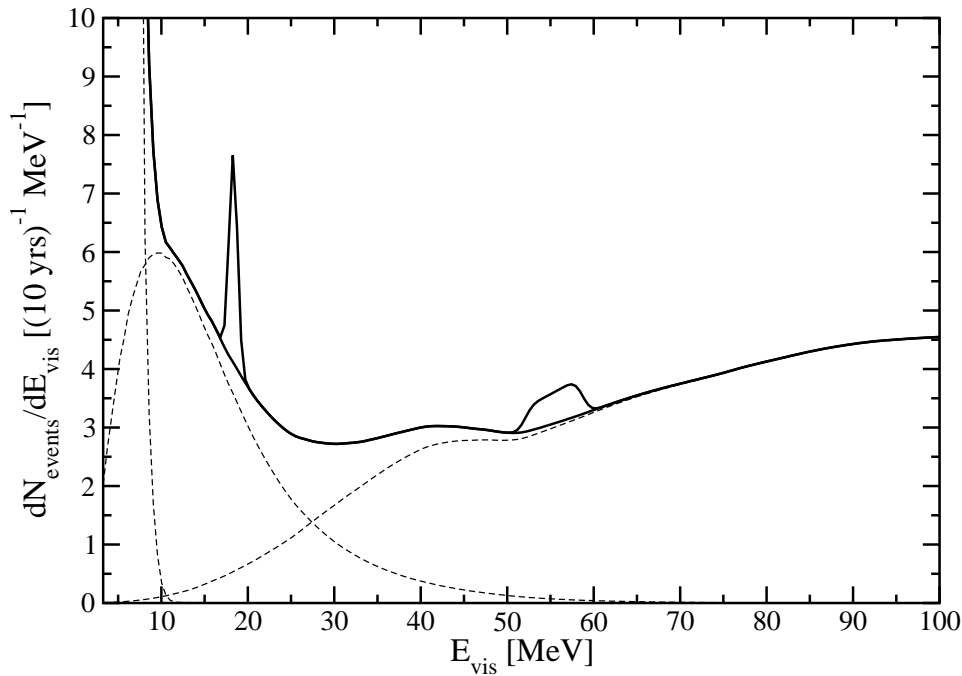}}
\caption{Expected signal in the proposed LENA detector, located in
  Pyh\"asalmi (Finland), after 10 years of running for two different
  values of the DM mass, $m_\chi = 20$~MeV and $m_\chi =
  60$~MeV. Dashed lines represent the individual contributions of each
  of the three different types of background events in this type of
  detector (reactor antineutrinos, DSNB and atmospheric neutrinos),
  whereas the solid lines represent the backgrounds plus the expected
  signal from DM annihilation in the Milky Way. We have used $\langle
  \sigma_{\rm A} v \rangle = 3 \times 10^{-26}$ cm$^3$ s$^{-1}$.}     
\label{LENA10years}
\end{figure} 

For the supernova rate per comoving volume we use the fit to
ultraviolet and far-infrared data obtained by Ref.~\cite{SFH} assuming
a modified Salpeter initial mass function with a turnover below 1
$M_\odot$~\cite{SalA}, and the parametric form for the star formation
rate of Ref.~\cite{2dFGRS},
\begin{equation}
\label{SNrate}
R_{\rm SN} (z) = 0.00915 M_\odot^{-1} \, \, \frac{(0.0119 + 0.091
  z)}{1+(z/3.3)^{5.3}}  ~.
\end{equation}
For the neutrino spectrum from each supernova, we consider the
simulation by the Lawrence Livermore group~\cite{LL} with the
parametrization for each flavor given by~\cite{KRJ03}
\begin{equation}
\label{SNispec}
\frac{dN_\nu}{dE_\nu} = \frac{(1+\beta_\nu)^{1+\beta_\nu} \,
  L_\nu}{\Gamma(1+\beta_\nu) \, \overline{E}_\nu^2} \,
  \left(\frac{E_\nu}{\overline{E}_\nu}\right)^{\beta_\nu} \,
  e^{-(1+\beta_\nu) E_\nu/\overline{E}_\nu} ~,
\end{equation}
with $\overline{E}_{\overline{\nu}_e} = 15.4$~MeV,
$\overline{E}_{\nu_x} = 21.6$~MeV, $\beta_{\overline{\nu}_e} = 3.8$,
$\beta_{\nu_x} = 1.8$, $L_{\overline{\nu}_e} = 4.9\times10^{52}$~ergs
and $L_{\nu_x} = 5.0\times10^{52}$~ergs~\cite{Ando04}, and where
$\nu_x$ represents non-electron-flavor neutrinos and
antineutrinos. Due to neutrino mixing, about 70\% ($|U_{e1}|^2 \simeq
0.7$) of the emitted $\overline{\nu}_e$ survive and 30\% ($1 -
|U_{e1}|^2 \simeq 0.3$) of the emitted $\nu_x$ will appear as
$\overline{\nu}_e$ at the Earth~\cite{Ando04}.

Finally, by plugging Eqs.~(\ref{SNrate})-(\ref{SNispec}) into
Eq.~(\ref{SNspec}), we obtain the expected DSNB flux at the Earth. Note
that the neutrino spectrum we have considered has a relatively large
average energy for each flavor (still consistent with star formation
history measurements~\cite{SFH}), which gives rise to a larger flux
than in other simulations~\cite{KRJ03,TBP03} at the energies where
the DSNB spectrum may dominate over other backgrounds ($\sim
10-30$~MeV). This is a conservative assumption for our purposes, for
this flux represents a background for a potential signal from DM
annihilation into neutrino-antineutrino pairs. Above $\sim$30~MeV, it
is the atmospheric $\overline{\nu}_e$ flux which gives the main
contribution to the background in a detector like LENA. The background
rate is calculated considering a gaussian energy resolution function
of width~\cite{LENADSNB}
\begin{equation}
\sigma_{\mathrm{LENA}} = 0.10 \, {\rm MeV} \, \sqrt{E/{\rm MeV}} ~.
\end{equation}
 
In Fig.~\ref{LENA10years}, we show the expected signal in the LENA
detector, if located in Pyh\"asalmi (Finland), along with the expected
backgrounds, after 10 years. We have considered a scintillator mixture
in weight of 20\% PXE (C$_{16}$H$_{18}$) and 80\% Dodecane
(C$_{12}$H$_{26}$) and a fiducial volume of 50$\times 10^3$ m$^3$,
which amounts to 3.3 $\times 10^{33}$ free protons. We depict the case
of two values for the DM mass, $m_\chi = 20$~MeV and $m_\chi =
60$~MeV, and we have considered here the canonical value for the halo
profile of Ref~\cite{YHBA}. In the case of low values of the masses,
e.g. $m_\chi =$ 20~MeV, even with the small rate predicted, a rather
easy discrimination between signal and background would be
possible. For higher values of the masses, in the few tens of MeV, as
clearly shown in Fig.~\ref{LENA10years}, the energy of the initial
neutrino cannot be uniquely reconstructed from the measured positron
energy. Therefore, the signal for $m_\chi = 60$~MeV is not
characterized by a delta-function in energy and has a spread over an
interval of $\sim$~10~MeV. In this case, the separation of the signal 
from the background will be more difficult, although still possible.

\section{Conclusions}
\label{conclusions}

In the present article we have considered the indirect bounds from the
annihilation into neutrino-antineutrino pairs of DM particles with
masses in the range of tens of MeV.  These particles have been
proposed as possible DM candidates and have been advocated in
different contexts to explain the 511~keV emission line from the bulge
as well as the observed values of neutrino masses. DM light particles
would self-annihilate in the galaxy producing a sizable flux of
neutrinos of energy $E_\nu = m_\chi$. These neutrinos might be detected
in present and future neutrino experiments. We have shown that the
present data from the SK experiment in the 18~MeV--82~MeV energy
region can constrain significantly the total DM annihilation cross
section to be smaller than $\sigma_{\rm A} \sim {\rm few} \times
(10^{-25}$--$10^{-26})\, {\rm cm}^3/{\rm s}$, depending on the
specific choice of the halo profile, for DM masses between $\sim$
15-100~MeV. We have also shown that future large neutrino detectors
with sufficient energy resolution and good background discrimination
could have the capability to observe a signal if these MeV particles
exist and the annihilation cross section is the one required to
reproduce the observed amount of DM. In particular, the LENA detector
would have the capability to find a positive signal in a large part of
the mass window under interest. A megaton-size water-\v{C}erenkov
detector, possibly doped with GdCl$_3$, and large Liquid Argon
detectors with very good energy resolution at MeV energies, would have
similar sensitivities. A negative signal in future large neutrino
experiments would either indicate that, if MeV DM exists, the
annihilation cross section at freeze-out was velocity-dependent or
exclude DM masses in the $\sim$~10--100~MeV. A positive signal would
imply that DM is constituted by particles with masses in the tens of
MeV range, would measure its mass and would determine, albeit halo
profile uncertainties, the cross section which was relevant at DM
freeze-out in the Early Universe.
 
\section*{Acknowledgments}
We are grateful to J.~Beacom for enlightening discussions and for a
careful reading of the manuscript. We also thank C.~Boehm and
H.~Yuksel for discussions, E.~Casanova-Plana for encouragement, and
M.~Wurm for clarifying some details about the LENA detector and for
providing us with the reactor antineutrino fluxes. SPR is partially
supported by the Spanish Grant FPA2005-01678 of the MCT.

\end{document}